\documentclass[runningheads]{svmult}

\usepackage{makeidx}   
\usepackage{graphicx}  
\usepackage{subeqnar}  
\usepackage{multicol}  
\usepackage{physprbb}  
\makeindex             

\begin{document}
\title*{Tracing the evolution of massive galaxies \protect\newline
up to z $\sim$ 3}
\toctitle{Tracing the evolution of massive galaxies
\protect\newline up to z $\sim$ 3}
%
%
\titlerunning{Tracing the evolution of massive galaxies}
%
\author{M. Longhetti \inst{1}
\and P. Saracco \inst{1}
\and S. Cristiani \inst{2}
\and A. Fontana \inst{3}
\and E. Giallongo \inst{3}
\and M. Nonino \inst{2}
\and E. Vanzella \inst{4}}

\authorrunning{M. Longhetti et al.}
%
%
\institute{INAF - Osservatorio Astronomico di Brera, Milano, Italy
\and European Southern Observatory
\and INAF - Osservatorio Astronomico di Roma, Italy
\and INAF - Osservatorio Astronomico di Trieste, Italy
}

\maketitle              


\section{J-K $>$ 3 galaxies: unveiling M$>10^{11}$M$_{\odot}$ 
early-type galaxies at z $\sim 3$}

Objects with unusual red near-IR color J-K$>$3 are extremely rare 
at magnitude brighter than K=20, while their surface density increases at fainter
magnitudes. The nature of these sources has not yet been firmly established
even if all the analysis performed so far conclude that they are galaxies
at z $>$ 2-3 ([1], [2], [3]).
A detailed analysis based on a multi-band data set 
(from 0.3$\mu$m to 2.15$\mu$m)
of three J-K$>$3 sources selected at Ks $<$ 22 
on the HDF-S ([4]) has shown that:
i) the three galaxies are at redshift 
2.5$<$z$<$3; ii) they have a stellar mass content of
M$_{star}\sim 10^{11}$M$_{\odot}$; iii) they would populate the bright end 
(L$_{z=0} \sim$ L$^{*}$) of the local luminosity function of galaxies 
even assuming they evolve passively; iv) they cannot follow an 
evolution significantly different from passive aging. 

\vskip -0.3 truecm
\section{Depicting the evolution of massive galaxies from z=3 to z=0}
Thus, these J-K$>$3 objects  selected are the z$\sim$3 counterpart 
of local massive early-type galaxies.
Between z=3 and z=0 we expect to find the same population of galaxies
at a different evolutionary status, namely with older stellar content. 
For istance, at z=1-1.5 we expect to find early-type galaxies (t$_{z=3}$ - 
t$_{z=1.5}$) $\approx$ 2 Gyr older. Such old early-type galaxies
have been found  among the EROs selected with R-K>5 (e.g.[5]).
In order to find their counterpart at the intermediate redshift z$\sim$2, 
we have identified a suitable color selection criterion by exploring the 
the expected colors evolution as derived by synthetic models ([6]). 
The results are displayed in Fig. 1: for short SF time scale ($\tau <$ 1Gyr 
galaxies at z$\sim$2 have colors J-K$<$3 and I-H$>$4.
In fact, adopting these selection criteria on the HDF-S data, we identified 4 massive 
evolved galaxies at 1.5$<z<$2.5 which seem to be the link between the J-K$>$3
galaxies (z$\sim3$) and the EROs (z$\sim$1). An example of this passive evolution 
connection is proposed in the right panel of Fig. 1:
822 (upper panel) a J-K$>$3 galaxy
selected in the HDF-S well described by a young SSP at z$\sim$2.9; 
1203 (middle panel) one of the four massive 
(M$_{star} \ge 10^{11}$ M$_{\odot}$) galaxies selected in the HDF-S on the basis 
of the selection criteria above described, well described
by a 1.4 Gyr old SSP at z$\sim$1.8; S7F5\_254
(lower panel) the spectrum of a massive early-type galaxy
at z=1.2 ([5]) superimposed on the template 
of a 3 Gyr old SSP.

\begin{figure}[b]
\vskip -0.5 truecm
\begin{center}
\includegraphics[width=0.8 \textwidth]{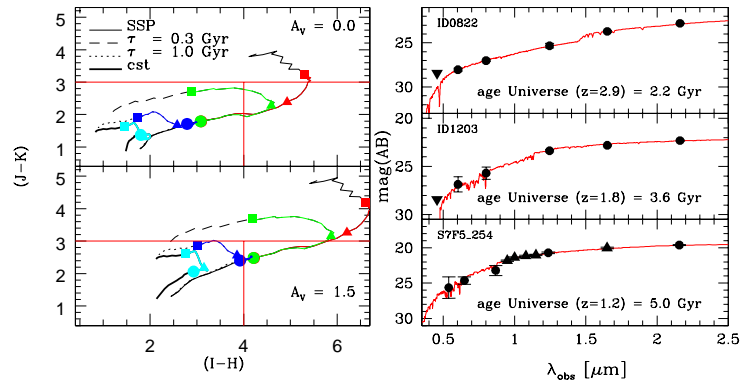}
\end{center}
\vskip -0.5 truecm
\caption[]{{\it left panel}: each line represents the evolution with redshift 
of the J-K and I-H colors for different SFHs between z=1 (filled circles), z=2
(filled triangles) and z=3 (filled squares)(solar metallicity, 
Salpeter IMF, formation redshift z$_{f}$=6); {\it right panel}: possible
passive evolution scenario (details in the text)}
\label{eps1}
\end{figure}

The photometry, the stellar masses and the mean ages of the stellar populations of 
these three massive galaxies at different redshift are apparently
those we expect from a massive galaxy fully assembled at z$\sim$3 which evolves 
passively in time.
This result is based, in fact, on a couple of objects and may be simply a coincidence. 
However, it is worthy to note that the number density of the three classes of objects 
represented in the figure and that of the local massive galaxies 
are well compatible with the depicted scenario. 
Actually, the number density of J-K$>$3 and K$<$22 galaxies 
(2.5$<$z$<$3, M$_{star} \ge 10^{11}$ M$_{\odot}$ galaxies) is 
$\rho=1.2 \times 10^{-4}$ Mpc$^{-3}$ ([4]) and that of 
J-K$<$3 and I-H$>$4 (1.5$<$z$<$2.5) calculated from the HDF-S data results to 
be the same within the errors. 
The number density of the early-type galaxies at 1.0$<$z$<$1.5 (R-K$>$5)
is consistent with the previous value once the same stellar
mass threshold (M$_{star} \ge 10^{11}$ M$_{\odot}$) is taken into account ([4]).
The local density of M$_{star} \ge 10^{11}$ M$_{\odot}$ elliptical galaxies
is $\rho \simeq 3 \times 10^{-4}$ Mpc$^{-3}$ ([7]): at least 40\% of the local
population of bright massive early-type galaxies can be explained as the results
of a simple passive evolution started from galaxies already assembled
at z$\sim$3.

%

\end{document}